\documentclass[12pt]{iopart}

\usepackage{graphicx,caption}
\usepackage[normalem]{ulem}
\begin{document}

\title[]{Valley and Zeeman Splittings in Multilayer Epitaxial Graphene Revealed by Circular Polarization Resolved Magneto-infrared Spectroscopy}

\author{Yuxuan Jiang$^1$, Zhengguang Lu$^{1,2}$, Jamey Gigliotti$^3$, Avinash Rustagi$^4$\footnote[7]{Current address: School of Electrical and Computer Engineering, Purdue University, West Lafayette, IN 47907}, Li Chen$^1$, Claire Berger$^{3,5}$, Walter A. de Heer$^{3,6}$, Christopher J. Stanton$^4$, Dmitry Smirnov$^1$ and Zhigang Jiang$^3$}

\address{$^1$National High Magnetic Field Laboratory, Tallahassee, Florida 32310, USA}
\address{$^2$Department of Physics, Florida State University, Tallahassee, Florida 32306, USA}
\address{$^3$School of Physics, Georgia Institute of Technology, Atlanta, Georgia 30332, USA}
\address{$^4$Department of Physics, University of Florida, Gainesville, Florida 32611, USA}
\address{$^5$Institut N\'{e}el, CNRS\textemdash Universit\'{e} Grenoble Alpes, 38042 Grenoble, France}
\address{$^6$Tianjin International Center of Nanoparticles and Nanosystems, Tianjin University, 300072 Tianjin, China}

\ead{zhigang.jiang@physics.gatech.edu}
\vspace{10pt}
\begin{abstract}
Circular polarization resolved magneto-infrared studies of multilayer epitaxial graphene (MEG) are performed using tunable quantum cascade lasers in high magnetic fields up to 17.5 T. Landau level (LL) transitions in the monolayer and bilayer graphene inclusions of MEG are resolved, and considerable electron-hole asymmetry is observed in the extracted electronic band structure. For monolayer graphene, a four-fold splitting of the $n=0$ to $n=1$ LL transition is evidenced and attributed to the lifting of the valley and spin degeneracy of the zeroth LL and the broken electron-hole symmetry. The magnetic field dependence of the splitting further reveals its possible mechanisms. The best fit to experimental data yields effective $g$-factors, $g^*_{VS}=6.7$ and $g^*_{ZS}=4.8$, for the valley and Zeeman splitting, respectively.
\end{abstract}


\maketitle

%

\section{Introduction}
Graphene has attracted great interests in the past 15 years due to its spectacular physical properties \cite{Rev_Beenakker,Rev_Geim,Rev_Peres,Rev_DasSarma,Rev_Goerbig,Rev_Neto,Rev_Basov}. The low-energy electronic structure of graphene features linearly dispersed conduction and valence bands, which touch at two inequivalent, charge-neutral points (namely $K$ and $K'$ points) in the Brillouin zone \cite{band_Wallace}. Therefore, electrons in graphene exhibit a four-fold degeneracy, accounting for the spin and $K$/$K'$ valley symmetry. Broken-symmetry states, particularly at the charge neutrality point of graphene, have long been a focal point of research \cite{Split_YZ,Split_ZJ,Split_Zhao,Split_AY,Split_AY_2}. To better resolve these states, high mobility graphene samples and high magnetic fields are typically required, enabling many-particle effects and enhanced spin (Zeeman) and valley splittings in energy. High-field and high-resolution spectroscopy are thus the preferred technique to probe the nature of the broken-symmetry states in graphene \cite{Split_JS,SU4_Orlita1,SU4_Orlita2,Split_Hunt,Split_He,Split_He_new}. It provides an accurate measure of the energy splittings as a function of magnetic field ($B$) for direct comparison with theory \cite{Split_KY}.

Epitaxial graphene grown on SiC \cite{deHeer1,Berger_deHeer} is an ideal platform for implementation of various spectroscopy techniques, owing to its large area, high transparency, and easy means of surface cleaning via high-temperature annealing. The multilayer epitaxial graphene (MEG) grown on the carbon-terminated face of SiC is particularly suited for studying the broken-symmetry states in charge-neutral graphene \cite{Split_JS,SU4_Orlita1,SU4_Orlita2}, as the top MEG layers are essentially decoupled from the polar surface of SiC \cite{Growth_Conrad2,First,Growth_Conrad3}, leading to a carrier density as low as $5\times 10^9$ cm$^{-2}$ \cite{UP_Orlita}. The rotational stacking between the top layers also makes them nearly decoupled from each other, rendering an electronic structure indistinguishable from that of an isolated monolayer graphene (MLG) \cite{Growth_Conrad3,UP_MP,Growth_Conrad1}. Record-high mobility of 250,000 cm$^2$/(V$\cdot$s) has been reported in top MEG layers, which remains constant at elevated temperatures even up to room temperature \cite{UP_Orlita}.

Magneto-infrared (magneto-IR) spectroscopy proves to be a useful tool in probing the broken-symmetry states in charge-neutral graphene \cite{SU4_Orlita1,SU4_Orlita2,Erik1,Chen1}. However, the direct evidence to date of the four-fold splitting of the zeroth Landau level (LL) only comes from the electronic transport and tunneling spectroscopy measurements, where the energy splittings are enhanced when the Fermi energy is placed within the gap between two split sub-LLs. The interpretation of the electronic transport measurements in Hall-bar-like device geometries is also complicated by the (extrinsic) conditions of the graphene edge \cite{Dean_Shen,Zhu_Geim}, which may be of a different nature from that in the bulk \cite{Dean}. In this work, via combining the bulk-sensitive circular-polarization (CP) resolved magneto-IR spectroscopy with high-quality quasi-neutral MEGs, we are able to probe the valley and Zeeman splittings (VS and ZS) of graphene LLs in high magnetic fields. CP-resolved measurements are known to be sensitive in revealing fine LL structures due to its selective activation of the electron-like or hole-like transitions \cite{CP_AJ,CP_MD,CP_VS}. Therefore, it can yield important information on the electron-hole asymmetry of the material's band structure. The long lifetime of Dirac fermions in MEG and the application of high magnetic fields also enable the high energy resolution of our measurements and distinguish our work from earlier CP-resolved studies of graphene \cite{CP_Kuz,CP_Kono}. Our magnetic field dependent measurements lead to the determination of effective $g$-factors for ZS and VS, which are key to understanding the rich phase diagram of charge-neutral graphene in high magnetic fields \cite{Phase_Kharitonov} and its implications in MEG.
\begin{figure*}[t!]
\includegraphics[width=\textwidth]{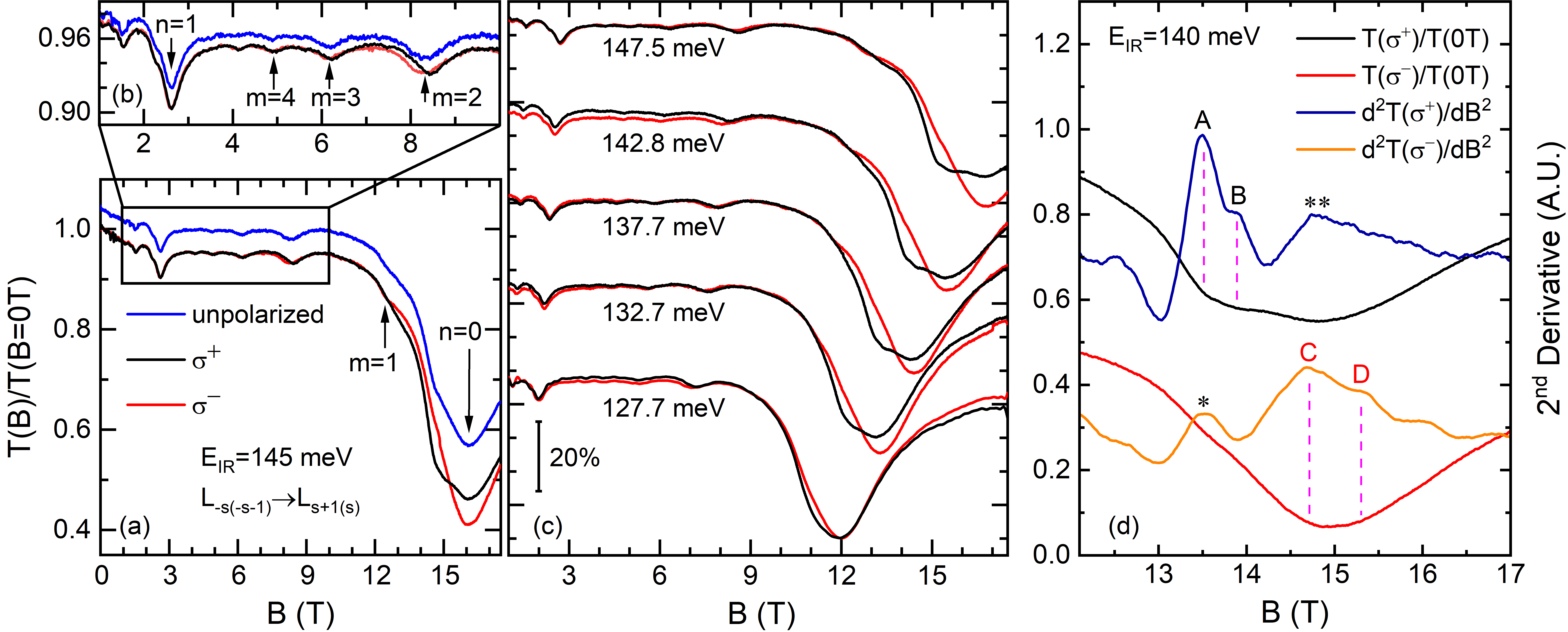}
\caption{(color online) (a) Normalized magneto-transmission spectra, $T(B)/T(B=0\ \rm{T})$, of MEG with $\sigma^+$ (black), $\sigma^-$ (red) and unpolarized (blue) incident IR light at the photon energy $E_{\rm{IR}}=145$ meV. The unpolarized spectrum is offset vertically for clarity. The allowed LL transitions in $\sigma^+$ ($\sigma^-$) polarized light are $\rm{LL}_{-s}$$\rightarrow$$\rm{LL}_{s+1}$ ($\rm{LL}_{-s-1}$$\rightarrow$$\rm{LL}_{s}$), where $s=|n|,|m|$ and integers $n$ and $m$ denote the LL indices of MLG and BLG, respectively. (b) Zoom-in view of the spectra in (a) between $B=1$ T and 10 T. (c) CP-resolved magneto-transmission spectra measured at different incident photon energies. (d) Comparison of the normalized magneto-transmission spectra with their second derivatives at the incident photon energy $E_{{\rm{IR}}}=140$ meV. The Roman letters label the four-fold splitting of the $n=0$ LL transition of MLG. The asterisk symbol ($\ast$) indicates a weak mode with minor spectral weight, likely due to the CP leakage of our setup, while the double asterisks ($\ast\ast$) label a broad mode that carries considerable spectral weight and with a transition energy corresponding to $v_F=1.00 \times 10^6$ m/s. In (c,d), the spectra are also offset vertically for clarity.}
\end{figure*}

\section{Method}
MEG samples were grown on the carbon-terminated face of SiC using the confinement controlled sublimation method \cite{deHeer1,Berger_deHeer}, followed by routine atomic force microscopy (Park System XE) and Raman spectroscopy characterizations. The high-quality samples were selected based on the surface morphology and the absence of the $D$ peak in Raman spectra \cite{Raman_Kunc}, and loaded in a home-built magneto-IR dipper equipped with a superconducting magnet at 4.2K. The IR transmission measurements were performed in Faraday configuration with both circularly polarized and unpolarized light emitted from a set of quantum cascade lasers covering the spectral range between 100 meV and 200 meV. However, due to the overlap with the SiC reststrahlen band, no transmission signal was detected below 124 meV. For CP-resolved measurements, the polarized light was generated by placing a linear polarizer and a wavelength-tunable quarter waveplate in the optical path. For consistency, the CP-resolved spectra were taken by fixing the light polarization and sweeping the magnetic field in positive or negative directions, which is equivalent to the use of $\sigma^+$ and $\sigma^-$ polarized light. The details of the experiment setup can be found in \cite{CP_YJ}.

\section{Results and discussions}
\subsection{Landau level spectra of multilayer epitaxial graphene}
Figure 1(a,b) show the typical magneto-transmission spectra of MEG, $T(B)/T(B=0\ \rm{T})$, measured with incident IR light at the photon energy $E_{\rm{IR}}=145$ meV and normalized to its zero field value. From the unpolarized (blue) spectra, one can identify a series of absorption dips or modes and, following previous work \cite{Bilayer_Orlita}, attribute them to two distinct sequences of inter-LL transitions, with the dominant sequence originated from MLG and the second sequence from AB-stacked bilayer graphene (BLG). The presence of a small amount of AB-stacked BLG inclusions in MEG is well understood from previous spectroscopy studies \cite{Bilayer_Orlita,Bilayer_Faugeras,Bilayer_Conrad,Bilayer_Lanzara}, that is, due to the stacking faults (AB-stack) in the otherwise rotationally stacked ($\sim$30$^\circ$ with respect to each other) and electronically decoupled top MEG layers. Therefore, all the absorption modes observed in our experiment can be described with the LL spectra of MLG and BLG \cite{IR_ZJ_1,IR_ZJ_2}
\begin{eqnarray}
E_{\mathrm{MLG},n}=&\mathrm{sgn}(n) \sqrt{2e\hbar v_F^2|n|B},\label{eq:MLG}\\
E_{\mathrm{BLG},m}=&\frac{\mathrm{sgn}(m)}{\sqrt{2}}[(2|m|+1)\Delta_B^2+\gamma_1^2\nonumber\\
&-\sqrt{\gamma_1^4+2(2|m|+1)\Delta_B^2\gamma_1^2+\Delta_B^4}\ ]^{\frac{1}{2}},\label{eq:BLG}
\end{eqnarray}
where $e$ is the elementary electron charge, $\hbar$ is the reduced Planck's constant, $v_F$ is the Fermi velocity, integer $n$ ($m$) is the LL index of MLG (BLG), and $n>0$ ($m>0$) or $n<0$ ($m<0$) represents electron or hole LLs. The Fermi velocity $v_F$ is also related to the intralayer hopping parameter $\gamma_0$ and determines the cyclotron energy of MLG through $\Delta_B\equiv v_F\sqrt{2e\hbar B}$. $\gamma_1$ is an interlayer coupling parameter of BLG, describing the relatively strong coupling between two carbon atoms stacked directly on top of each other. The corresponding LL transitions are expected to be $\rm{LL}_{-s(-s-1)}$$\rightarrow$$\rm{LL}_{s+1(s)}$ (which are referred to as the $s$ transition throughout this work), following the usual selection rule $\Delta s=\pm 1$ with integer $s=|n|,|m|$.

Using $\sigma^+$ and $\sigma^-$ polarized light, one can perform the CP-resolved measurements and probe the electron-like ($\rm{LL}_{-s}$$\rightarrow$$\rm{LL}_{s+1}$, $\Delta s=+1$, $\sigma^+$ active) and hole-like ($\rm{LL}_{-s-1}$$\rightarrow$$\rm{LL}_{s}$, $\Delta s=-1$, $\sigma^-$ active) transitions separately. Figure 1(a,b,c) show the CP-resolved magneto-transmission spectra of MEG measured at selected incident photon energies. A prominent difference in spectral lineshape between the $\rm{LL}_{n=0}$$\rightarrow$$\rm{LL}_{n=1}$ (black) and $\rm{LL}_{n=-1}$$\rightarrow$$\rm{LL}_{n=0}$ (red) transitions is clearly evidenced, and it becomes more pronounced with increasing photon energy. This observation is suggestive of considerable electron-hole asymmetry in the electronic structure of top MEG layers, which, in the literature \cite{eh_Nicholas}, has simply been described by assigning a larger Fermi velocity for the electrons than the holes, $v_F^e>v_F^h$.

It is worth noting that asymmetry in magneto-transmission spectra has previously been reported in the CP-resolved measurements of a \textit{thin} MEG sample \cite{CP_Kuz}, where a low energy mode is observed only with $\sigma^+$ polarized light and attributed to the $\rm{LL}_{n=0}$$\rightarrow$$\rm{LL}_{n=1}$ transition in the doped layers close to the graphene/SiC interface. The doped graphene layers near the interface are known to have low electron mobility and thus exhibit a smaller Fermi velocity and lower LL transition energies for a given magnetic field \cite{Fertig,Martinez}. In our case, however, the additional $\sigma^+$-active mode(s) occur at a lower magnetic field (figure 1(a,c)), that is, higher in energy if measured in a constant magnetic field, as compared to the $\sigma^-$-active mode(s). This is in sharp contrast to \cite{CP_Kuz}. In addition, our MEG sample is much thicker, similar to that studied in \cite{UP_Orlita} with record-high mobility.

\subsection{Four-fold splitting of the $n=0$ Landau level transition}
Further careful inspection and second derivative calculation of the measured magneto-transmission spectra of MEG reveal fine structures or modes within the $n=0$ LL transition, as shown in figure 1(d). Four distinct modes, labeled by Roman letters A, B, C, and D, can be readily identified via correlating a kink feature in the spectra with a peak in its second derivative, especially at high incident photon energies. The additional peak, labeled by the asterisk symbol ($\ast$) in second derivative, indicates a weak $\sigma^-$-active mode with minor spectral weight, likely due to the CP leakage of our experiment setup. The double asterisks ($\ast\ast$), on the other hand, label a broad $\sigma^+$-active mode that carries considerable spectral weight and with a transition energy corresponding to $v_F=1.00 \times 10^6$ m/s. Since in high mobility MLG, the many-particle effects tend to renormalize the bands, giving rise to a larger Fermi velocity near the charge neutrality \cite{IR_ZJ_1,vF_Li}, and cause ZS and VS of the zeroth LL, one can attribute the A, B, C, and D modes to the four-fold splitting of the $n=0$ transition in high mobility graphene layers, whereas attribute the $\ast\ast$ mode to the layers with (relatively) low mobility.
\begin{figure}[t!]
\centering
\includegraphics[width=0.6\textwidth]{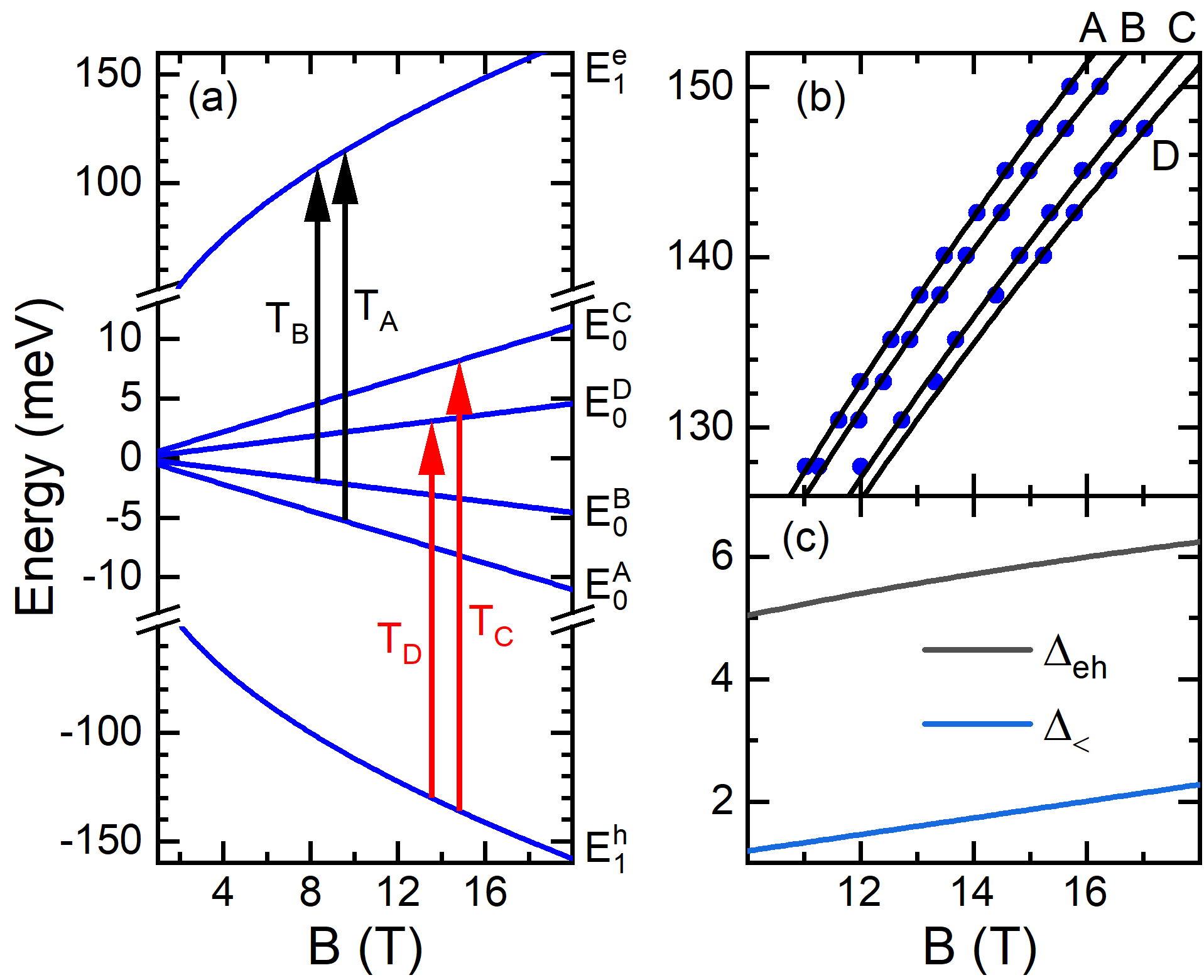}
\caption{(color online) (a) Landau fan diagram of MLG near charge neutrality. The four-fold splitting of the $n=0$ transition is labeled by $T_{\rm A}$, $T_{\rm B}$, $T_{\rm C}$, and $T_{\rm D}$, corresponding to the mode $A$, $B$, $C$, and $D$ in figure 1(d), respectively. Since the top MEG layers are quasi-neutral, none of these transitions are expected to be Pauli-blocked. (b) Evolution of the mode $A$, $B$, $C$, and $D$ (blue dots) in the energy--magnetic-field space. The solid lines are best fit to the data using equation \eref{interpolate}. (c) Magnetic field dependence of $\Delta_<$ and $\Delta_{eh}$ calculated from equation \eref{Delta}.}
\end{figure}

Figure 2(a) shows the Landau fan diagram of MLG near charge neutrality, considering the four-fold splitting of the zeroth LL and the electron-hole asymmetry. For simplicity, we schematically neglect the splittings of the first electron and hole LLs, $E_1^e$ and $E_1^h$, which can simply be added back in at the end of our analysis. The modes labeled by $A$, $B$, $C$, and $D$ in figure 1(d) can then be assigned to $T_{\rm A}$, $T_{\rm B}$, $T_{\rm C}$, and $T_{\rm D}$ transitions in figure 2(a), given that $T_{\rm A}$ and $T_{\rm B}$ are electron-like and $T_{\rm A}>T_{\rm B}$ while $T_{\rm C}$ and $T_{\rm D}$ are hole-like and $T_{\rm C}>T_{\rm D}$. These transitions can be quantitatively described using the following LL or sub-LL energies:
\begin{equation}
\label{LL}
\eqalign{
E_1^{e}&=v_F^e \sqrt{2e\hbar B},\quad E_1^{h}=-v_F^h \sqrt{2e\hbar B}, \\
E_0^{\mathrm C}&=-E_0^{\mathrm A}=\frac{\Delta_>(B)+\Delta_<(B)}{2},\\
E_0^{\mathrm D}&=-E_0^{\mathrm B}=\frac{\Delta_>(B)-\Delta_<(B)}{2},
}
\end{equation}
where the electron-hole asymmetry is reflected in $E_1^e$ and $E_1^h$, and since there is no consensus on the relative magnitude of the ZS and VS, $\Delta_>$ and $\Delta_<$ are employed to denote the larger and smaller splitting between the two, respectively. Both $\Delta_>$ and $\Delta_<$ are expected to have a distinct magnetic field dependence, depending on its underlying mechanism to be discussed below.

From equation \eref{LL}, one can deduce the values of $\Delta_<$ and $\Delta_{eh}\equiv E_1^e+E_1^h$ (which describes the degree of electron-hole asymmetry at a given magnetic field) using experimental parameters $T_{\rm A,B,C,D}$
\begin{equation}
\label{Delta}
\eqalign{
\Delta_<=\frac{(T_{\mathrm A}-T_{\mathrm B})+(T_{\mathrm C}-T_{\mathrm D})}{2},\\
\Delta_{eh}=\frac{(T_{\mathrm A}+T_{\mathrm B})-(T_{\mathrm C}+T_{\mathrm D})}{2}.
}
\end{equation}
Practically, this is done by taking the CP-resolved measurements at various incident photon energies and extracting the corresponding magnetic fields of $T_{\rm A,B,C,D}$ transitions. The results are then plotted in the energy--magnetic-field space (as shown in figure 2(b)), where the energy of each mode can be interpolated for any given magnetic field. For best interpolation, one can fit the magnetic field dependence of each mode with
\begin{eqnarray}
\label{interpolate}
T_i=a_i\sqrt{B}+b_i B, \hspace{1cm} i={\mathrm A,B,C,D}
\end{eqnarray}
as the VS and ZS of the zeroth LL in MLG are expected to be either $\propto$$\sqrt{B}$ or $\propto$$B$ \cite{Phase_Kharitonov}. In equation \eref{interpolate}, $a_i$ and $b_i$ are the fitting parameters for mode $i$.

Figure 2(c) shows the deduced $\Delta_<$ and $\Delta_{eh}$ as a function of magnetic field. Although the effective magnetic field range of the measurements (for the $n=0$ LL transition) is limited to between 11 and 17 T, one can still ascertain a linear-in-$B$ dependence of $\Delta_<$ with $\Delta_<=0$ meV at zero magnetic field. Such a linear dependence helps identify $\Delta_<$ as Zeeman-like splitting with an enhanced $g$-factor while leaving $\Delta_>$ as a result of broken valley degeneracy of the zeroth LL. This assignment excludes the spin-polarized ferromagnetic state as the ground state of charge-neutral graphene, in accordance with prior studies \cite{Split_AY,Split_AY_2,SU4_Orlita1,SU4_Orlita2,Split_He_new}.

In addition, as a consistency check, the magnetic field dependence of $\Delta_{eh}$ (figure 2(c)) is examined and found to be $\propto$$\sqrt{B}$, considering $\Delta_{eh}=0$ meV as $B\rightarrow 0$. This observation is consistent with that expected from equation \eref{LL}, validating the interpolation procedure undertaken. The magnitude of $\Delta_{eh}$, however, is several times larger than $\Delta_<$ and comparable with $\Delta_>$ (to be discussed next). Therefore, electron-hole asymmetry is an important factor in determining the splitting energies.

\subsection{Valley and Zeeman splittings of the zeroth Landau level in monolayer graphene}
Except for the spin-polarized ferromagnetic state, there are three other possible symmetry-breaking states theoretically predicted for $\Delta_>$ of MLG \cite{Phase_Kharitonov}, namely, the canted antiferromagnetic, charge density wave and Kekul\'{e}-distortion states. Although the previous electronic transport study \cite{Split_AY_2} of MLG encapsulated in h-BN is in favor of the canted antiferromagnetic state (where the edge states dominate charge transport), bulk-sensitive spectroscopy studies of MEG lead to the charge density wave \cite{SU4_Orlita1,SU4_Orlita2} and Kekul\'{e}-distortion \cite{Split_He_new} state interpretations, leaving the nature of $\Delta_>$ still an open question. Unfortunately, since the determination of $\Delta_>$ requires additional information about the Fermi velocity than just the experimental parameters $T_{\rm A,B,C,D}$ and our model describing the electron-hole asymmetry is oversimplified, this work cannot give a conclusive answer to this question. Instead, we carry out a validity check on the two possible scenarios, (i) $\Delta_>\propto\sqrt{B}$ \cite{Split_KY,Shovkovy} and (ii) $\Delta_>\propto B$ \cite{Fuchs}, and provides a set of parameters that can best explain the experimental data.
\begin{figure}[t!]
\centering
\includegraphics[width=0.6\textwidth]{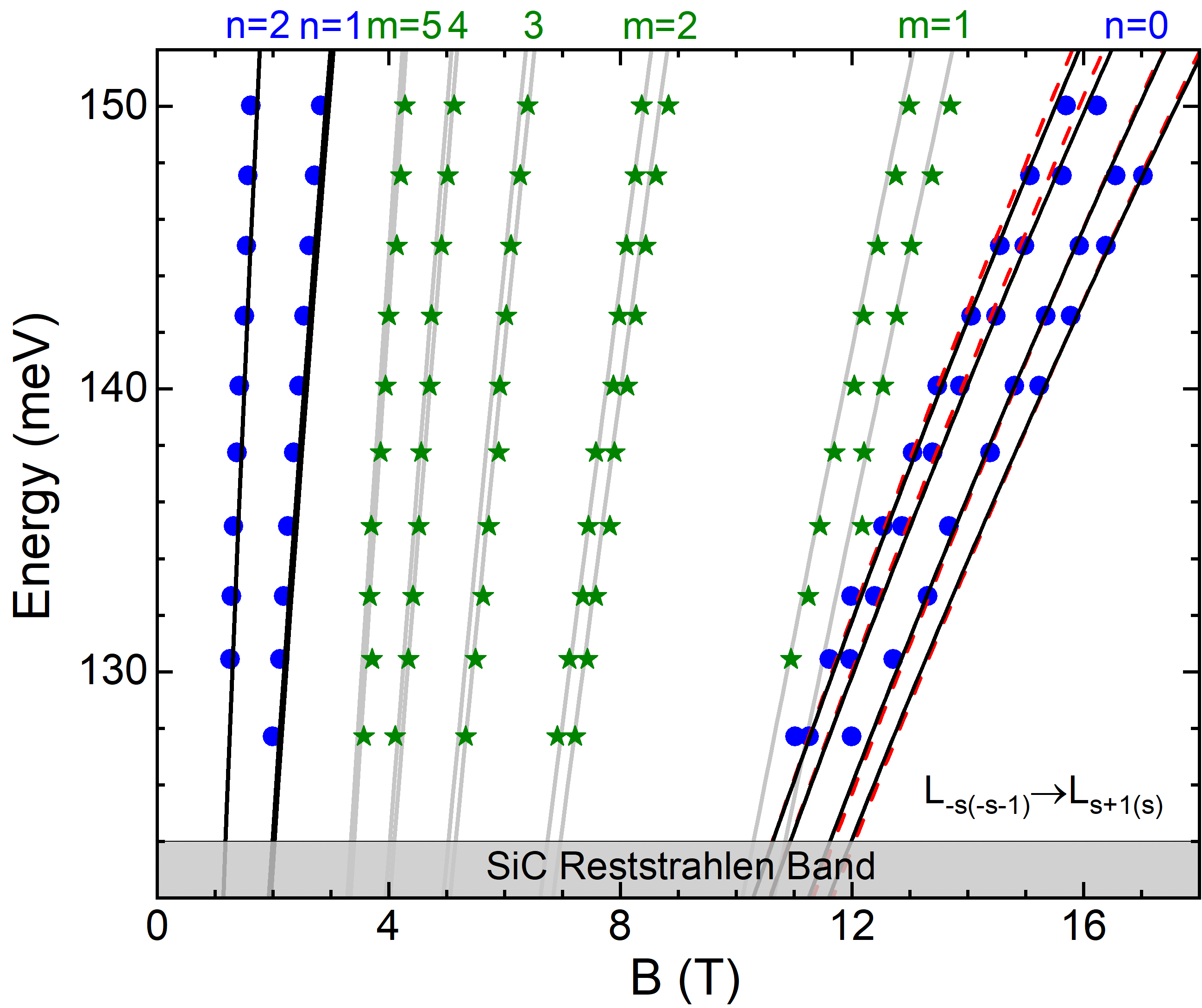}
\caption{(color online) Magnetic field dependence of the observed LL transitions from the MLG (blue dots) and BLG (green stars) inclusions in MEG. The black solid (red dash) lines are the best fit to the MLG transitions using equations \eref{eq:MLG} and \eref{LL} and assuming $\Delta_>\propto\sqrt{B}$ ($\Delta_> \propto B$), while the gray lines are the best fit to the BLG transitions. The darker gray area represents the experimentally inaccessible spectral region due to the SiC reststrahlen band.}
\end{figure}

To determine $\Delta_>$, one can fit all the LL transitions (MLG) observed, including the four-fold splitting of the $n=0$ transition, the $n=1$ and $n=2$ transitions with equations \eref{eq:MLG} and \eref{LL}. The black solid lines in figure 3 show the best fit to the data with $v_F^e=1.025 \times 10^6$ m/s, $v_F^h=0.975 \times 10^6$ m/s, $\Delta_<=0.16$ meV/T, and $\Delta_>=1.44$ meV/$\sqrt{\rm T}$ for case (i). The fit for case (ii) only results in small differences in the $n=0$ transition, as indicated by the red dash lines in figure 3 with $\Delta_>=0.39$ meV/T. Although within the experimental uncertainty, one cannot differentiate the above two cases, it is still insightful to have a closer look at the fitting parameters. First, the extracted Fermi velocities and the electron-hole asymmetry ($\pm2.5\%$) are consistent with those reported in previous works \cite{First,UP_Orlita,UP_MP,eh_Nicholas,vF_Li}. Second, $\Delta_<=0.16$ meV/T is corresponding to a ZS with a $g$-factor of 2.8. After considering the additional contribution from the ZS of the first LL with a bare electron $g$-factor of 2 \cite{Split_YZ}, the total effective $g$-factor reaches $g^*_{ZS}=4.8$. Third, $\Delta_>=1.44$ meV/$\sqrt{\rm T}$ of case (i) can be attributed to an electron-electron interaction induced VS. However, the induced energy gap is estimated to be on the order of Coulomb energy \cite{Split_KY}, $e^2 /4\pi \epsilon \epsilon_0 l_B\approx 11$ meV/$\sqrt{\rm T}$, much larger than the experimental value. Here, $l_B=\sqrt{\hbar/eB}$ is the magnetic length and $\epsilon\approx 5$ and $\epsilon_0$ are the relative permittivity of MEG and the vacuum permittivity, respectively \cite{PL_ZJ}. Lastly, $\Delta_>=0.39$ meV/T of case (ii) can either originate from the charge density wave \cite{SU4_Orlita1,SU4_Orlita2} or Kekul\'{e}-distortion \cite{Split_He_new} state as a result of the electron-phonon interaction. Such interaction lowers the ground state energy via slightly distorting the graphene lattice in a way that breaks the inversion symmetry and leads to VS \cite{Fuchs}. The resulting energy contribution is proportional to the electron degeneracy of each sub-LL, therefore exhibiting a linear-in-$B$ dependence. This case seems to best explain our experimental findings, with an effective VS $g$-factor of $g^*_{VS}=6.7$.

\subsection{Electron-hole asymmetry anomaly in bilayer graphene}
The CP-resolved spectra of the BLG inclusions in MEG reveal yet another interesting behavior, that is, pronounced electron-hole asymmetry but with the opposite sign to that in MLG. Figure 4(a) shows the normalized magneto-transmission spectra of the $m=2$ LL transition of BLG measured with $\sigma^+$ (black) and $\sigma^-$ (red) polarized light at different incident photon energies. Here, the $\sigma^-$-active transitions (dips) always appear on the low field side of the $\sigma^+$-active transitions, in sharp contrast to the $n=0$ transition of MLG in figure 1(c). Such behavior occurs in all the BLG transitions observed, especially the $m=1$ and $m=2$ transitions, as also implied in figure 3.
\begin{figure}[t!]
\centering
\includegraphics[width=0.6\textwidth]{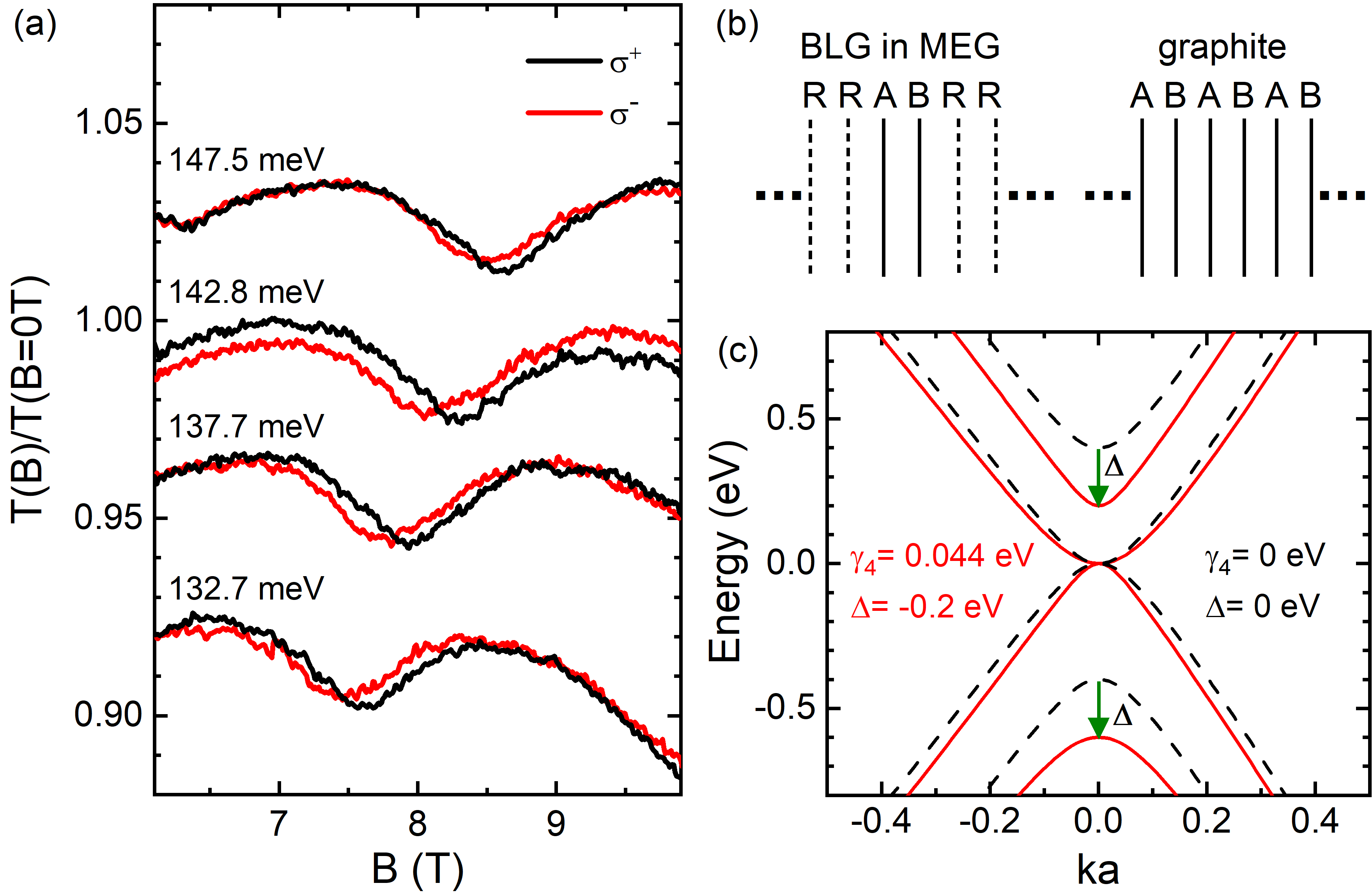}
\caption{(color online) (a) Normalized magneto-transmission spectra of the $m=2$ LL transition of BLG measured with $\sigma^+$ (black) and $\sigma^-$ (red) polarized light at different incident photon energies. (b) Schematics of the rotational stacking order between graphene layers in MEG and AB-stacked graphite. The letters A, B, and R denote three different rotation angles of a graphene plane, which are 0$^\circ$, 60$^\circ$, and 30$^\circ$, respectively. (c) Low-energy band structure of BLG with (red) and without (black) electron-hole asymmetry. For both cases, the bands are plotted using $\gamma_0=3.16$ eV and $\gamma_1=0.4$ eV. $k$ is the wave vector and $a=2.46$ $\rm{\AA}$ is the lattice constant of graphene. The electron-hole asymmetry is introduced by assigning $\gamma_4=0.044$ eV and $\Delta=-0.2$ eV. For demonstration purpose, $\Delta$ is taken to be about three times larger than the extracted value ($-0.068$ eV) from experiment.}
\end{figure}

To quantitatively understand this behavior, one can fit the BLG transitions (green stars in figure 3) with equation \eref{eq:BLG}. Due to the massive Dirac fermion nature of the electrons and holes in BLG, the fitting can be performed by fixing the Fermi velocity to its MLG value, $v_F=1.02 \times 10^6$ m/s, measured with unpolarized light \cite{Bilayer_Orlita}, whereas differentiating the band mass of electrons and holes via $m^*_{e,h}=\gamma_1^{e,h}/2v_F^2$. The gray lines in figure 3 show the best fit to the five BLG transitions observed in the experiment, and the corresponding fitting parameters are $m^*_e=0.0376 m_0$ and $m^*_h=0.0283 m_0$, where $m_0$ is the bare electron mass. The magnitude of the electron-hole asymmetry is then $\pm 14\%$, consistent with that extracted from the splittings of the $m=1$ and $m=2$ transitions in the previous magneto-IR study of MEG \cite{Bilayer_Orlita}.

The sign of the electron-hole asymmetry in the BLG inclusions, however, is in stark contrast to the existing literature on exfoliated BLG \cite{IR_ZJ_2,BLGEH_Malard,BLGEH_Zhang,BLGEH_Li,BLGEH_Fai}, where a lighter electron mass is always expected. Such an anomaly ought to be associated with the multilayer nature of MEG, particularly the coupling between the BLG and its neighboring layers. To examine this possibility, one can take an extreme case and compare the tight-binding parameters of BLG with AB-stacked graphite \cite{M_review}. The presence of additional graphene layers in graphite is found to strongly influence the skew interlayer coupling parameter $\gamma_4$ and the energy difference $\Delta$ between the dimer and non-dimer sites. Since $\gamma_4$ and $\Delta$ are the two primary sources for the electron-hole asymmetry in BLG and the value of $\Delta$ can be either positive or negative \cite{M_review,Falko_review}, it is not surprising to see a different sign for the electron-hole asymmetry in MEG due to its unique stacking order (figure 4(b)) and coupling between layers. Figure 4(c) shows a possible electronic band structure of BLG that can explain our experimental data. Here, instead of using the phenomenological band mass $m^*_{e,h}$, the electron-hole asymmetry is introduced by $\gamma_4$ and $\Delta$, while setting $\gamma_0=3.16$ eV (corresponding to $v_F=1.02 \times 10^6$ m/s) and $\gamma_1=0.4$ eV. $m^*_e>m^*_h$ observed in the experiment corresponds to $\Delta<0$, which could be due to the local potential changes caused by the rotational stacked graphene layers (with different rotation angles) above and below the BLG and the next-nearest layer couplings \cite{M_review,Falko_review}. Quantitatively, the amount of electron-hole asymmetry can be expressed as $\pm(\frac{\Delta}{\gamma_1}+\frac{2\gamma_4}{\gamma_0})=\pm 14\%$ \cite{Bilayer_Orlita,M_review}. By fixing $\gamma_4=0.044$ eV, as that in graphite, one can obtain $\Delta=-0.068$ eV for MEG.

\section{Conclusion}
In conclusion, we have performed a magneto-IR spectroscopy study of high-quality MEG with tunable CP light. We find that the MLG inclusions in MEG feature a four-fold splitting of the $n=0$ LL transition, resulting from the lifting of the valley and spin degeneracy of the zeroth LL and the broken electron-hole symmetry. By analyzing the magnetic field dependence of the transition energies, we deduce a possible scenario that involves VS at the charge neutrality and enhanced ZS in the electron and hole sub-LLs. The extracted effective $g$-factors are $g^*_{VS}=6.7$ and $g^*_{ZS}=4.8$, respectively. The CP-resolved measurements of the BLG inclusions uncover an even larger electron-hole asymmetry, with an opposite sign to the MLG. We show that the asymmetry could be strongly influenced by the stacking orientation of the BLG (with respect to the neighboring layers), making it a possible design parameter for future epitaxial graphene band engineering.

\ack{}
We thank Edward Conrad, Phillip First, Markus Kindermann, and Kun Yang for helpful discussions. This work was primarily supported by the DOE (Grant No. DE-FG02-07ER46451). The MEG growth and characterization at GT were supported by the NSF (Grant No. DMR-0820382) and the NASA Solar System Exploration Research Virtual Institute (cooperative agreement NNA17BF68A). The magneto-IR measurements were performed at the NHMFL, which is supported by the NSF through cooperative agreement DMR-1157490/1644779 and the State of Florida. A.R. and C.J.S. were supported by the Air Force Office of Scientific Research under Award Nos. FA9550-14-1-0376 and FA9550-17-1-0341. L.C. was supported in part by the DOE, Office of BES through Grant No. DE-SC0002140. C.B. acknowledges partial funding from the EU flagship graphene (Grant No. 604391). Z.J. acknowledges support from the NHMFL Visiting Scientist Program.

\end{document}